\def\aap{\ {A\&A}\ }

\def\apj{\ {ApJ}\ }
\def\apjl{\ {ApJL}\ }
\def\apjs{\ {ApJS}\ }

\def\mnras{\ {MNRAS}\ }

\def\pasj{\ {Publ. Astr. Soc. Japan}\ }

\def\ergs{{\rm\thinspace ergs}}

\documentclass[11pt,twoside]{article}

\usepackage{asp2010}
\usepackage{epsfig}
\usepackage{graphicx}
\usepackage{verbatim}
\usepackage{moreverb}

\usepackage{lscape}
\usepackage{natbib}
\usepackage{epsf}
\usepackage{graphicx}
\newcommand{\Tcr}{\mbox{$t_{\rm hc}$}}

\def\apgt{\ {\raise-.5ex\hbox{$\buildrel>\over\sim$}}\ } 
\def\aplt{\ {\raise-.5ex\hbox{$\buildrel<\over\sim$}}\ } 
\def\lt{\ {\raise-.5ex\hbox{$\buildrel>$}}\ } 
\def\gt{\ {\raise-.5ex\hbox{$\buildrel<$}}\ }

\def\Msun{\ensuremath{{\rm M}_{\odot}}}

\begin{document}
%\runauthor{Portegies Zwart at al.}
%\begin{frontmatter}

\title{Multi-physics simulations using a hierarchical interchangeable
  software interface}

\author{Simon Portegies Zwart,$^{1}$
        Stephen McMillan,$^{2}$
        Inti Pelupessy,$^{1}$
        Arjen van Elteren$^{1}$}

\affil{
  $^{1}$ Sterrewacht Leiden, P.O. Box 9513, 2300 RA Leiden,
	The Netherlands\\ 
$^{2}$ Department of Physics, Drexel University, Philadelphia, PA
	19104, USA 
}

\begin{abstract}
  We introduce a general-purpose framework for interconnecting
  scientific simulation programs using a homogeneous, unified software
  interface.  Our framework is intrinsically parallel, and
  conveniently separates all components in memory. It performs unit 
  conversion between different modules automatically and defines common data
  structures to communicate across different codes.  We use the framework 
  to simulate embedded star clusters.  For this purpose we couple solvers 
  for gravitational
  dynamics, stellar evolution and hydrodynamics to self consistently
  resolve the dynamical evolution simultaneousy with the internal
  nuclear evolution of the stars and the hydrodynamic response of the
  gas.  We find, in contrast to earlier studies, that the survival of
  a young star cluster depends only weakly on the efficiency of star
  formation.  The main reason for this weak dependency is the
  asymmetric expulsion of the embedding gas from the cluster.
\end{abstract}
%\end{frontmatter}

\section{Introduction}\label{Sect:Introduction}

Large-scale, high-resolution computer simulations dominate many areas
of theoretical and computational astrophysics.  The demand for such
simulations has expanded steadily over the past decade, and is likely
to continue to grow in coming years due to the relentless increase in
the volume, precision, and dynamic range of experimental data, as well
as the ever-widening spectral coverage of observations.  In order to
accommodate the improved observations, simulations must see a
comparable improvement in detail.  In recent years, simulation
environments have grown substantially by incorporating more detailed
descriptions of more physical processes, but the fundamental design of
the underlying codes has remained unchanged since the introduction of
object-oriented programming \cite{1962LISP.book..109M} and patterns
\cite{Patterns1987}.  As a result, maintaining and extending existing
large-scale, multi-physics solvers has become a major undertaking.
The legacy of design choices made long ago can hamper further
code development and expansion, prevent scaling on large parallel
computers, and render maintenance almost impossible.

The root cause of the increase in code complexity lies in the
traditional approach of incorporating multi-physics components into a
single simulation---namely, solving the equations appropriate to all
components in a monolithic software suite, often written by a
single researcher.  This monolithic solution may seem desirable from
the standpoint of consistency and performance, but the resulting
software generally suffers from fundamental problems in maintenance
and expansion.

The AMUSE (Astrophysical MUltipurpose Software
Environment)\footnote{see {\tt www.amusecdoe.org}.} project
assimilates well tested applications into a software suite
with which we can perform individual tasks, or reassemble the parts
into a new application that combines a wide variety of solvers. The
interfaces of codes within a common domain are designed to be as 
homogeneous as possible. This approach is possible in
astrophysics because of the tradition among astronomers during the
last several decades of sharing scientific software.  Many of these
applications were written by experts who spent careers developing
these codes and using them to conduct a wide range of numerical
experiments.  These packages are generally developed and maintained
independently of one another.  We refer to them collectively as
``community'' software.  AMUSE has recently surpassed our ``Noah's
ark'' developmental milestone
\citep{2009NewA...14..369P}, in which we have at least two numerical
solvers for each of the astrophysical domains of interest:
gravitational dynamics, stellar evolution, hydrodynamics, and
radiative transfer.

In a first step towards using this framework, we combine here three of
these fundamental ingredients of AMUSE to address a long-standing
problem in astrophysics: the relevance of the star formation
efficiency to the survival of embedded star clusters.

\section{AMUSE}

The AMUSE environment allows astrophysical codes from different
domains to be combined to conduct numerical experiments.  The
community codes are generally written independently, so AMUSE
encompasses a wide variety of computer languages and programming
styles.  The fundamental design feature of the framework is the
abstraction of the functionality of individual community codes behind
physically motivated interfaces that hide their complexity and/or
numerical implementation.  AMUSE presents the user with standard
building blocks that can be combined into applications and numerical
experiments.

The binding language that stitches the codes together is Python.  The
relatively low speed of this high-level language is not an issue,
since the focus in the high-level management code is not so much
performance (the computational cost being concentrated in the
component codes), but algorithmic flexibility and ease of programming
to allow rapid prototyping.  As described in more detail in the
contribution by McMillan, Portegies Zwart, and van Elteren elsewhere
in these proceedings, an AMUSE application consists of a Python user
script controlling one or more community modules.  The user script
specifies the initial conditions, selects the simulation modules, and
manages their use.  The coupling between the user script and a
community code is handled by the community module, which contains an
MPI-based communication interface onto the code, as well as
unit-handling facilities and an object oriented data model.

The relationships among the community codes define the numerical
experiment.  Our model here combines the effects of the self-gravity
and nuclear evolution of the stars with the hydrodynamics of the
intracluster gas (Pelupessy \& Portegies Zwart 2011, in preparation).
The latter includes both the primordial gas content of the cluster and
the gas liberated by the stars via stellar winds and supernovae.  In
this case we construct a hybrid N-body/stellar/hydrodynamic solver by
combining a direct N-body integrator, a stellar evolution package, and
an SPH code.

The Python user script controlling the experiment generates the
initial conditions (masses, positions and velocities of the stars, and
the distribution of the gas), specifies the various solvers,
structures the procedural calling sequences, resolves all interactions
among the various physical domains (e.g. feedback from the stellar
winds and supernovae to the surrounding gas), and processes the
output.  The particular modules employed in this experiment are the
Gadget-2 SPH code \cite{2005MNRAS.364.1105S}, the PhiGRAPE Hermite
N-body code \cite{2007NewA...12..357H}, the tree gravity code {\tt
Octgrav}
\cite{2010arXiv1005.5384G} and the
stellar evolution code \cite{2000MNRAS.315..543H}.  The first
dynamical model is used for the integration of the equations of motion
of the stars, the second module is used for the gravitational coupling
between the gas-particles and the stars. The combined solver consists
of an integrator for the coupled gas/gravitational dynamics systems
and a feedback prescription for mechanical energy input from the
evolving stars.

The gas and gravitational dynamics are coupled via the BRIDGE
integrator \citep{2007PASJ...59.1095F}.  BRIDGE provides a
semi-symplectic mapping for gravitational evolution in cases where the
dynamics of a system can be split into two (or more) distinct regimes.
A typical application would be a dense star cluster in a galaxy, where
the internal dynamics of the former evolves on a relatively short
timescale compared to the dynamics of the latter.  A similar idea was
implemented by \cite{2010PASJ...62..301S} by splitting the
gravitational and hydrodynamic evolution operators for simulating
gas-rich galaxy mergers.  They expressed the algorithm in a single
monolithic code, whereas we adopt the concept of operator splitting
within AMUSE to couple different codes.

\section{Initial conditions}

The clusters we simulate are composed of a mixture of gas and $N =
1000$ stars; both are distributed in a \cite{1911MNRAS..71..460P}
sphere, and they have the same characteristic radius. Stellar masses
are assigned using a \cite{1955ApJ...121..161S} IMF between 0.1 and
100\,\Msun, with an additional constraint that the most massive star
is $\sim 22$\,\Msun. This maximum mass is based on the most massive
star naively expected for a cluster with this number of stars and mass
function \cite{2003ApJ...598.1076K}. The masses of the stars are
assigned independently of their positions in the cluster.  We present
here the results of two of our simulations, which in our larger paper
describing this work are identified as model A2 and model A5 (see
Pelupessy \& Portegies Zwart 2011). 

For small clusters the number of high-mass stars can vary quite
substantially between different realizations of the IMF, and we have
performed simulations of models A2 and A5 with numerous random
realizations of the IMF to examine this effect.  We have performed
additional simulations in which we varied the number of gas particles
to test whether our results are independent of the resolution of the
gas dynamics.

\section{Results}

\begin{figure}[htbp]
\begin{center}
\psfig{figure=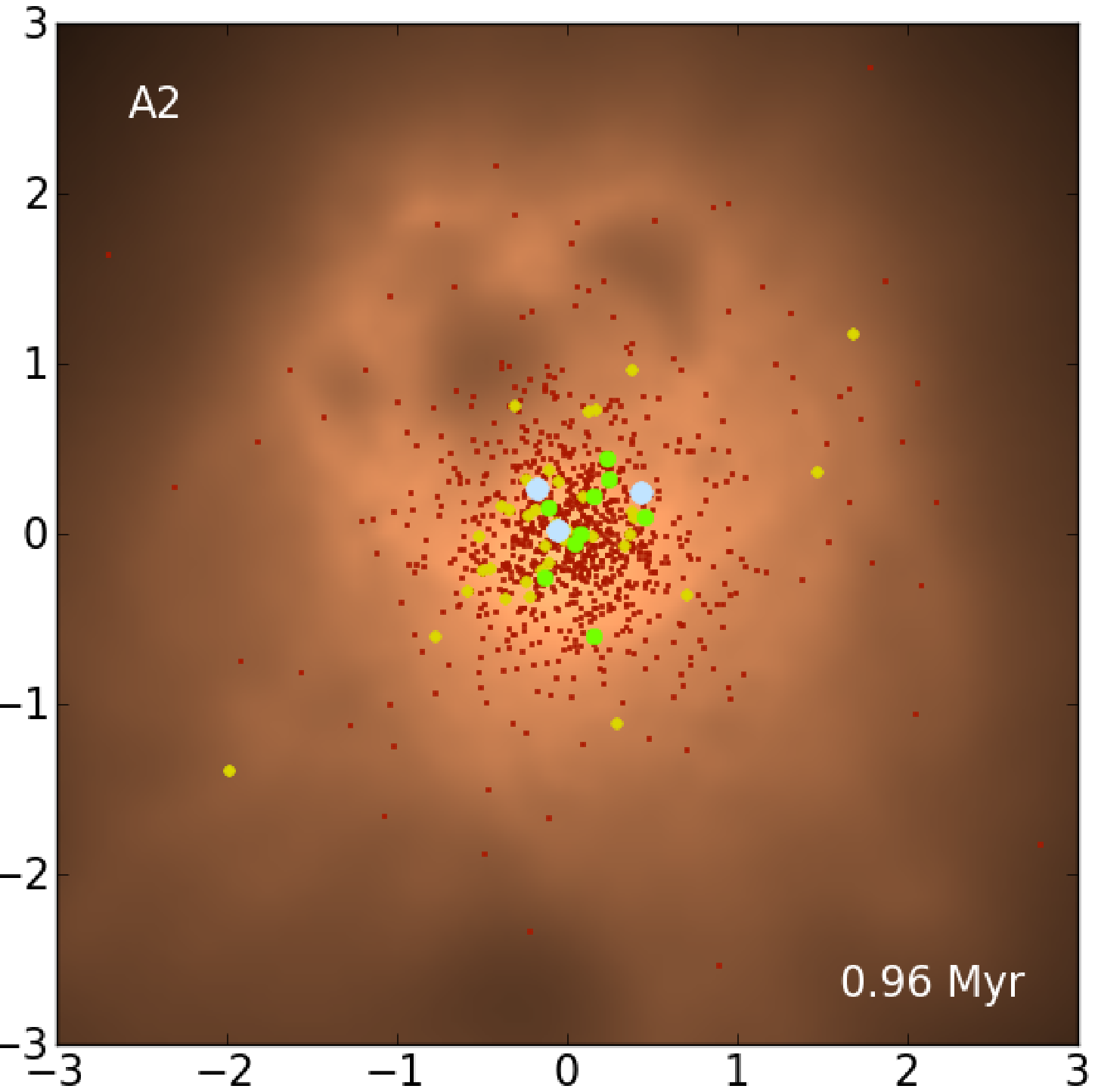,height=.3\textwidth} %,angle=-90} 
\psfig{figure=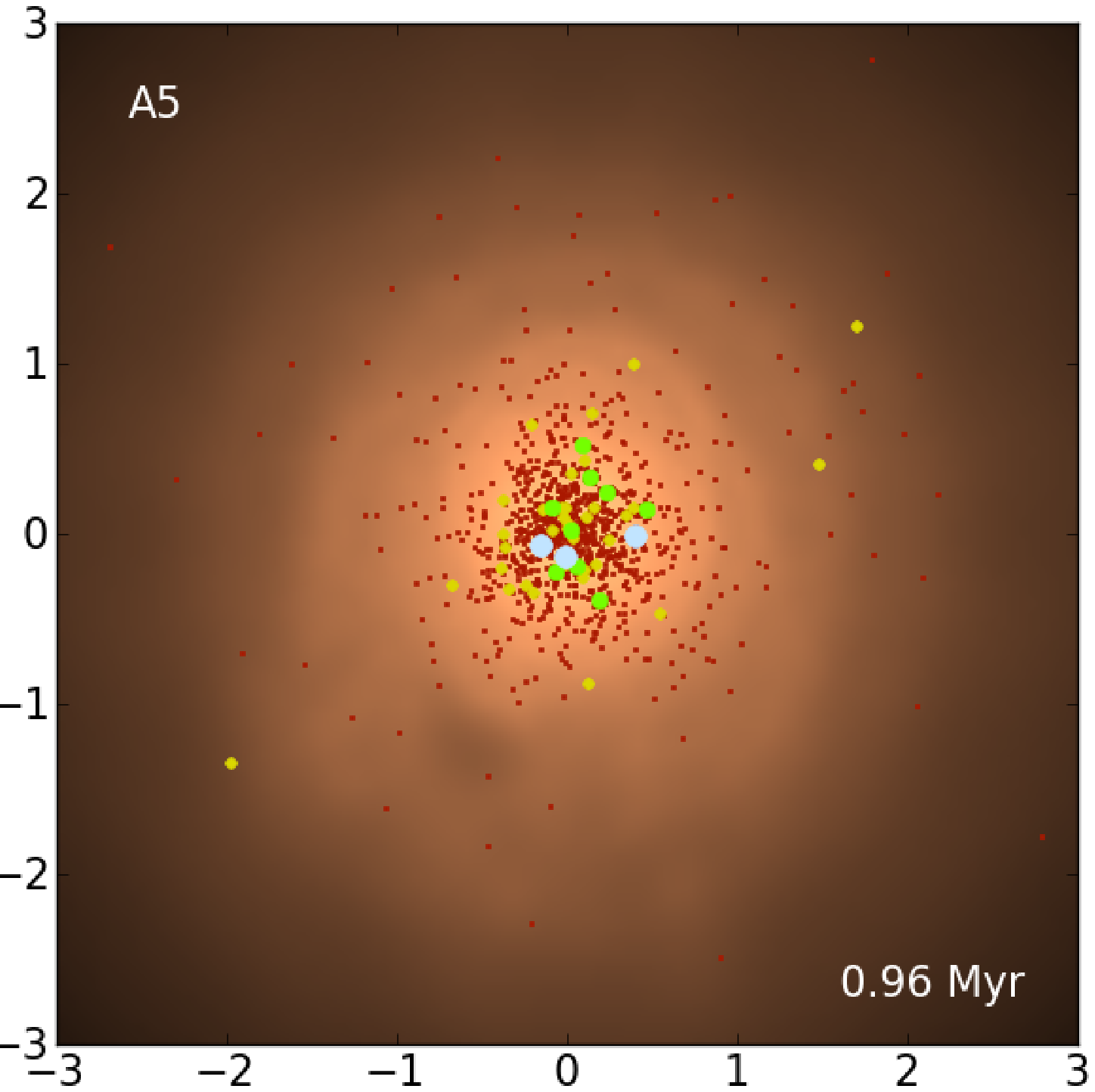,height=.3\textwidth} %,angle=-90} 

\psfig{figure=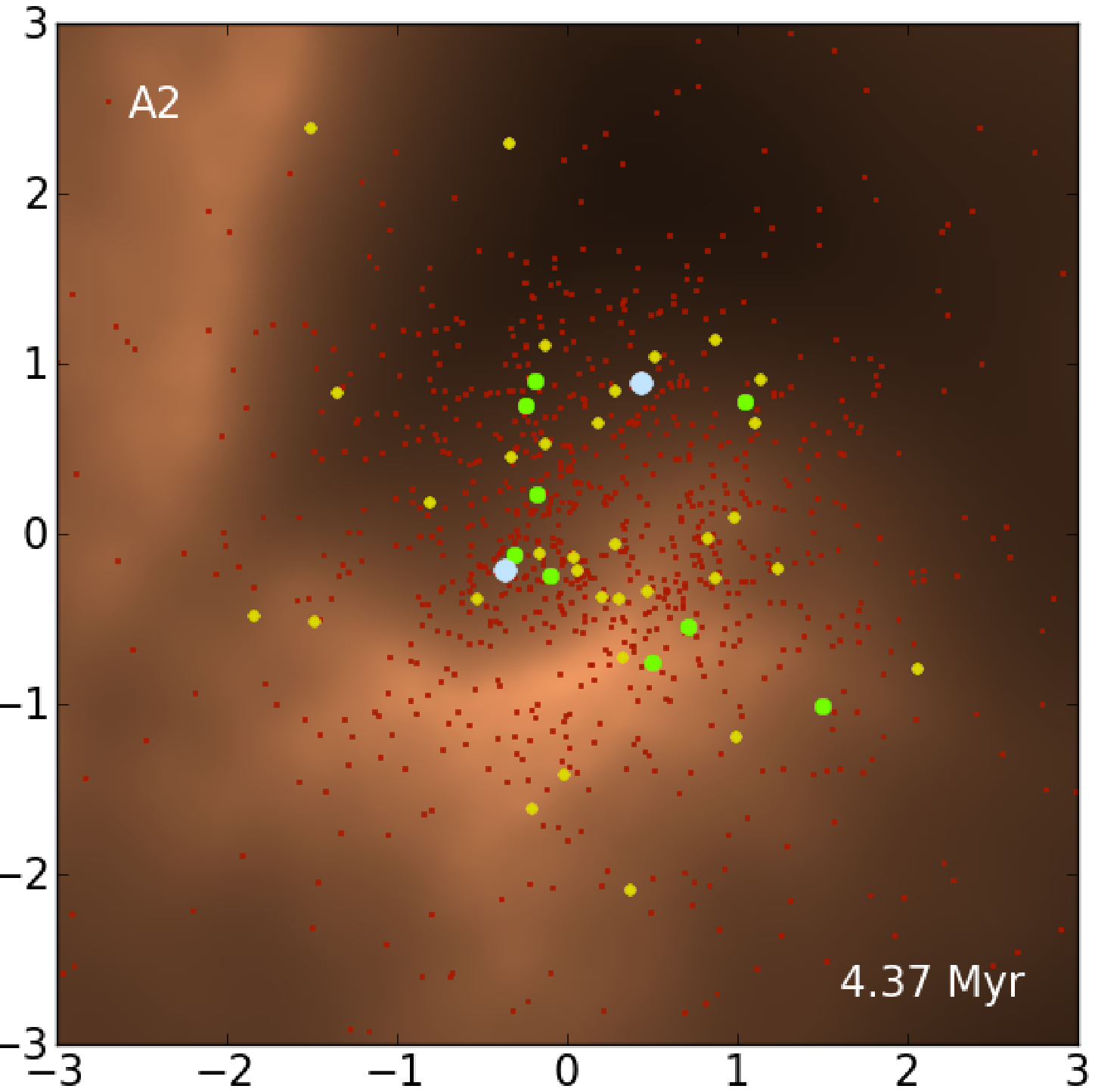,height=.3\textwidth} %,angle=-90} 
\psfig{figure=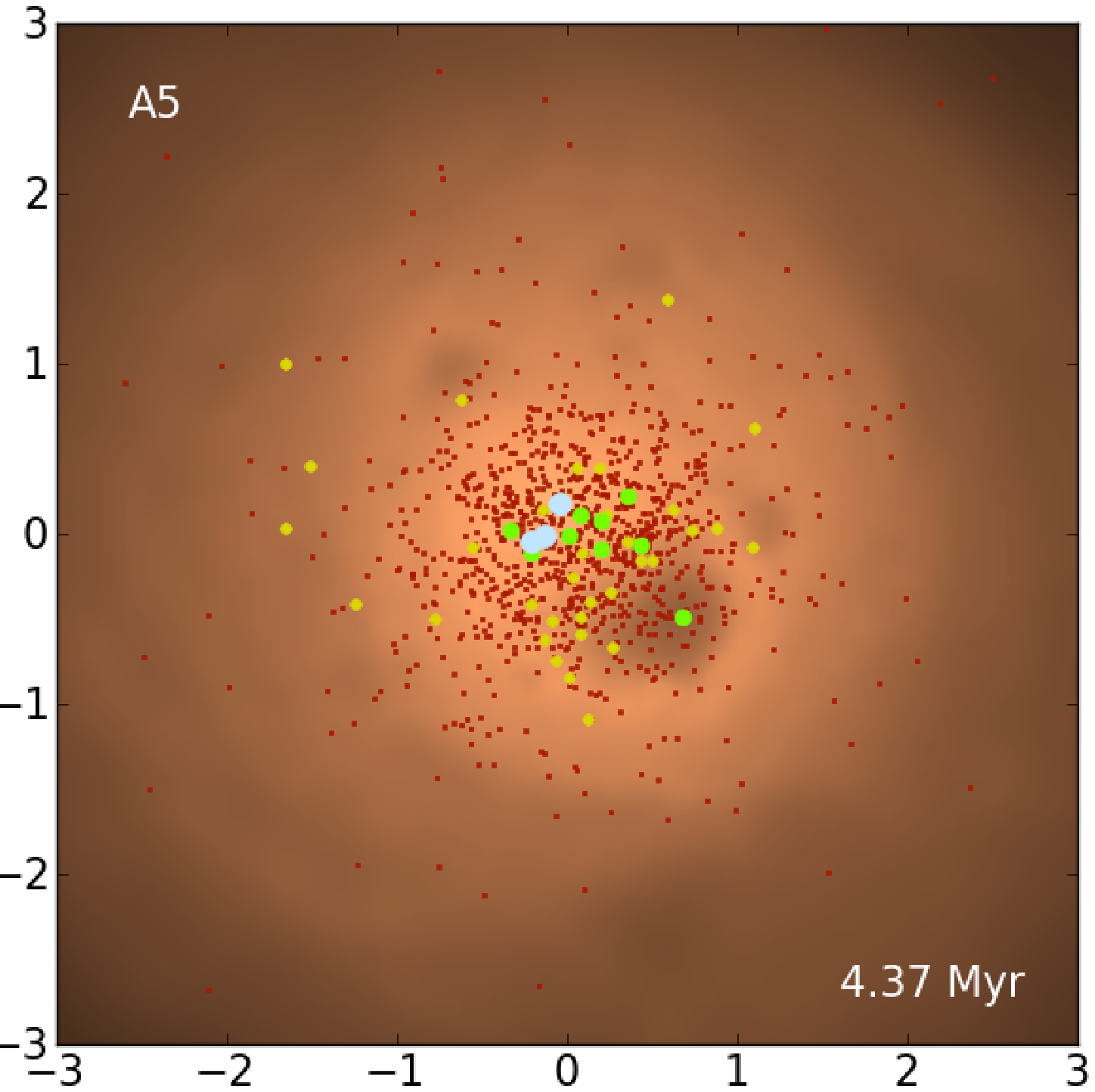,height=.3\textwidth} %,angle=-90} 

\psfig{figure=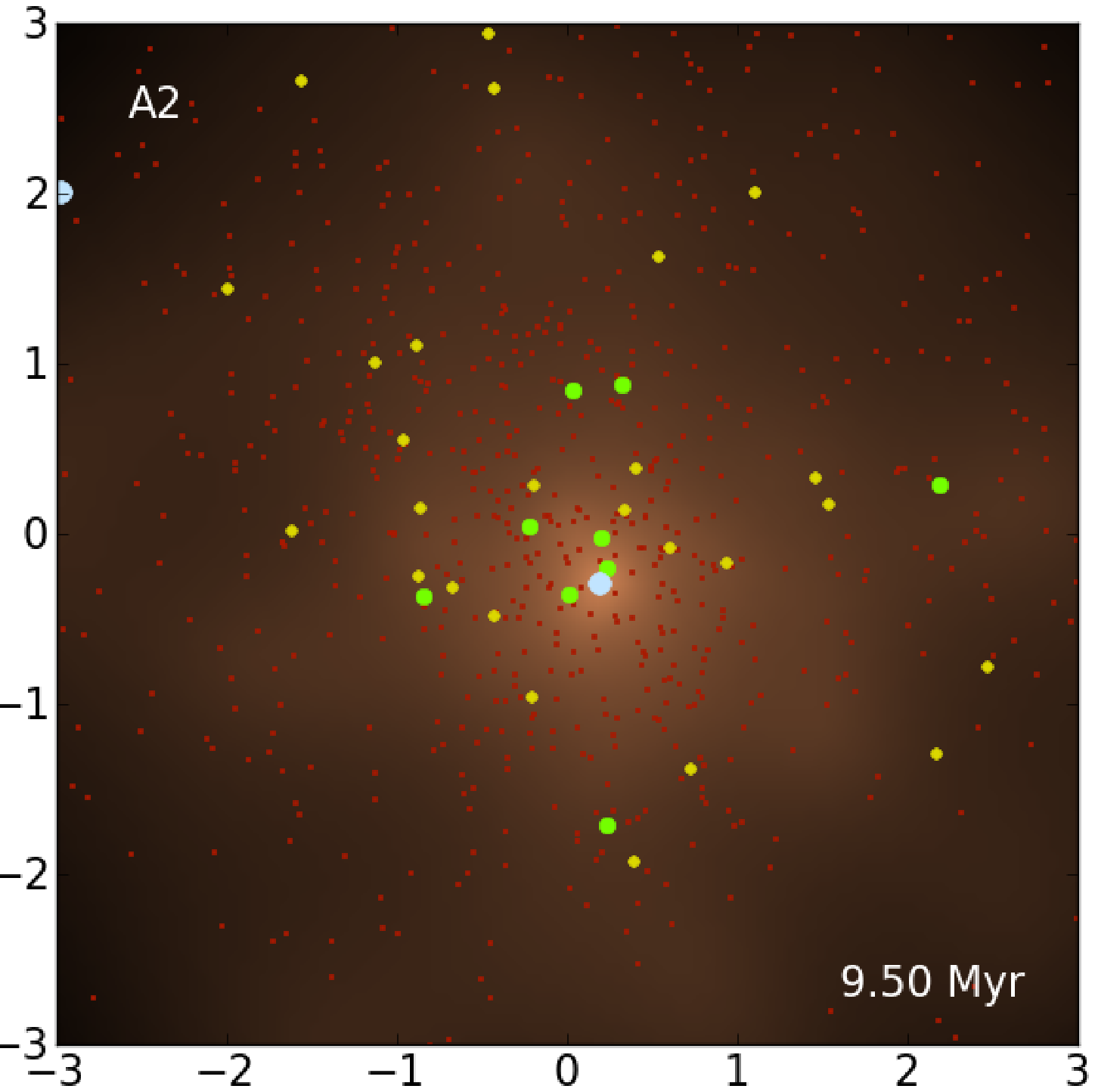,height=.3\textwidth} %,angle=-90} 
\psfig{figure=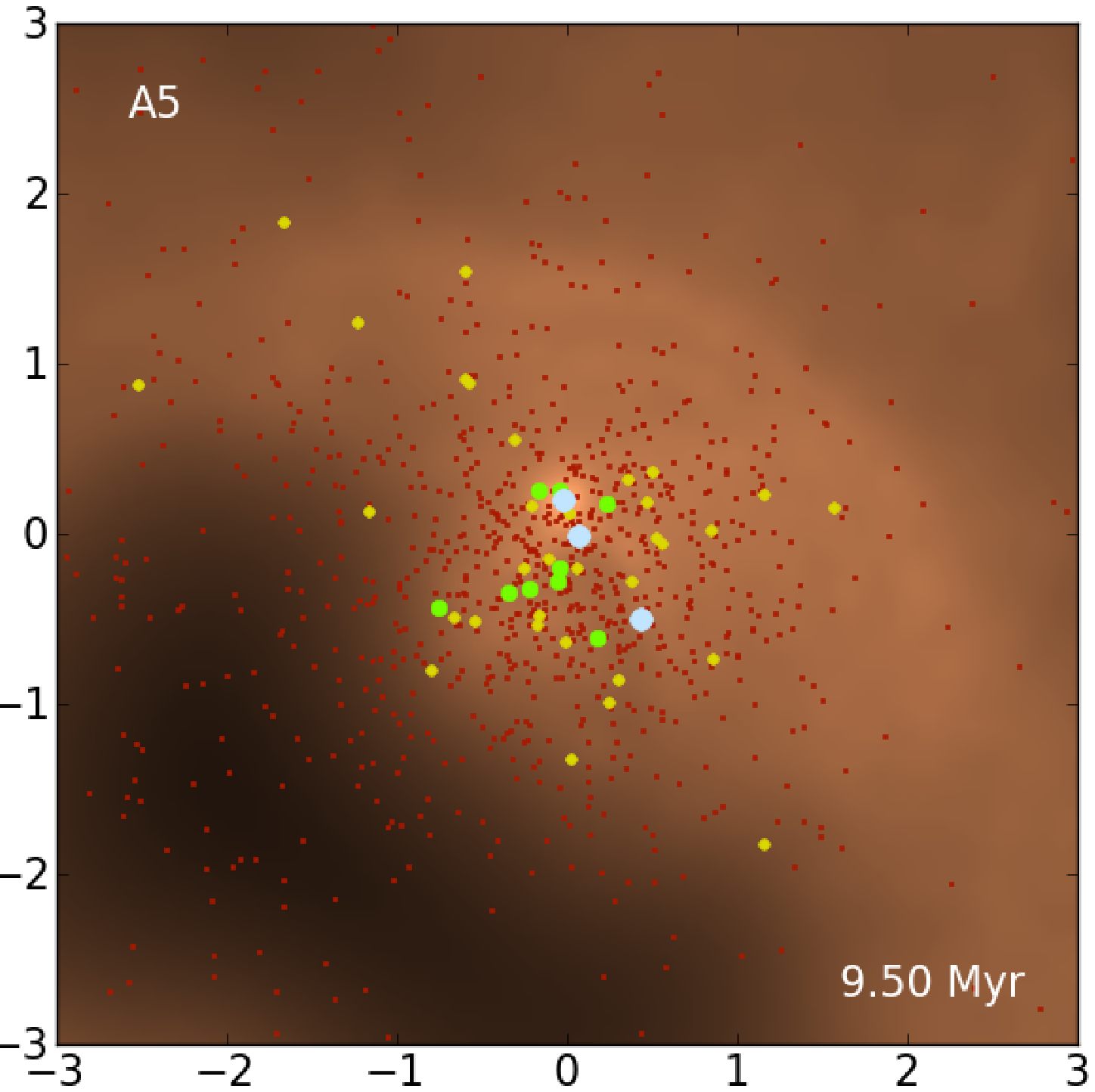,height=.3\textwidth} %,angle=-90} 

\psfig{figure=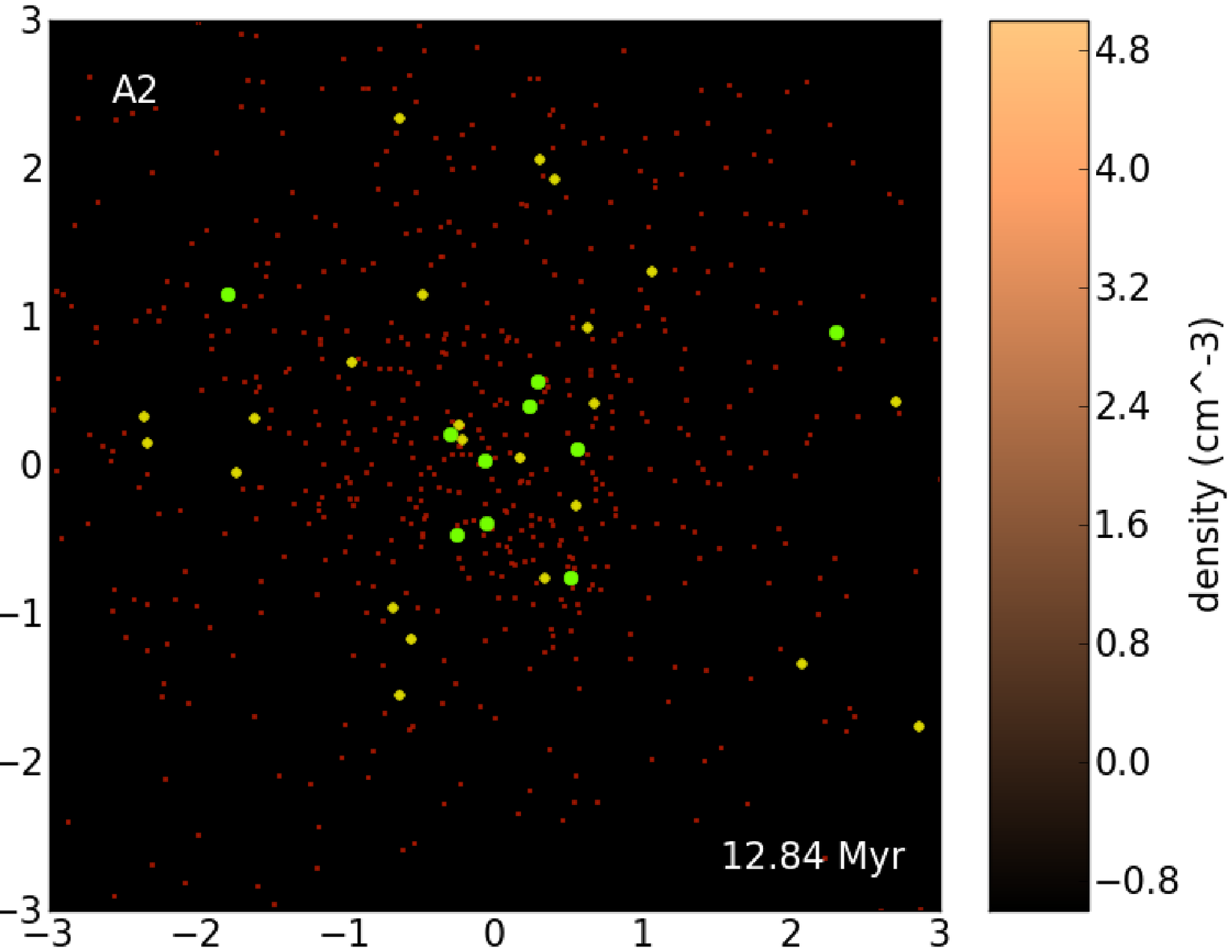,height=.3\textwidth}  %,angle=-90} 
\psfig{figure=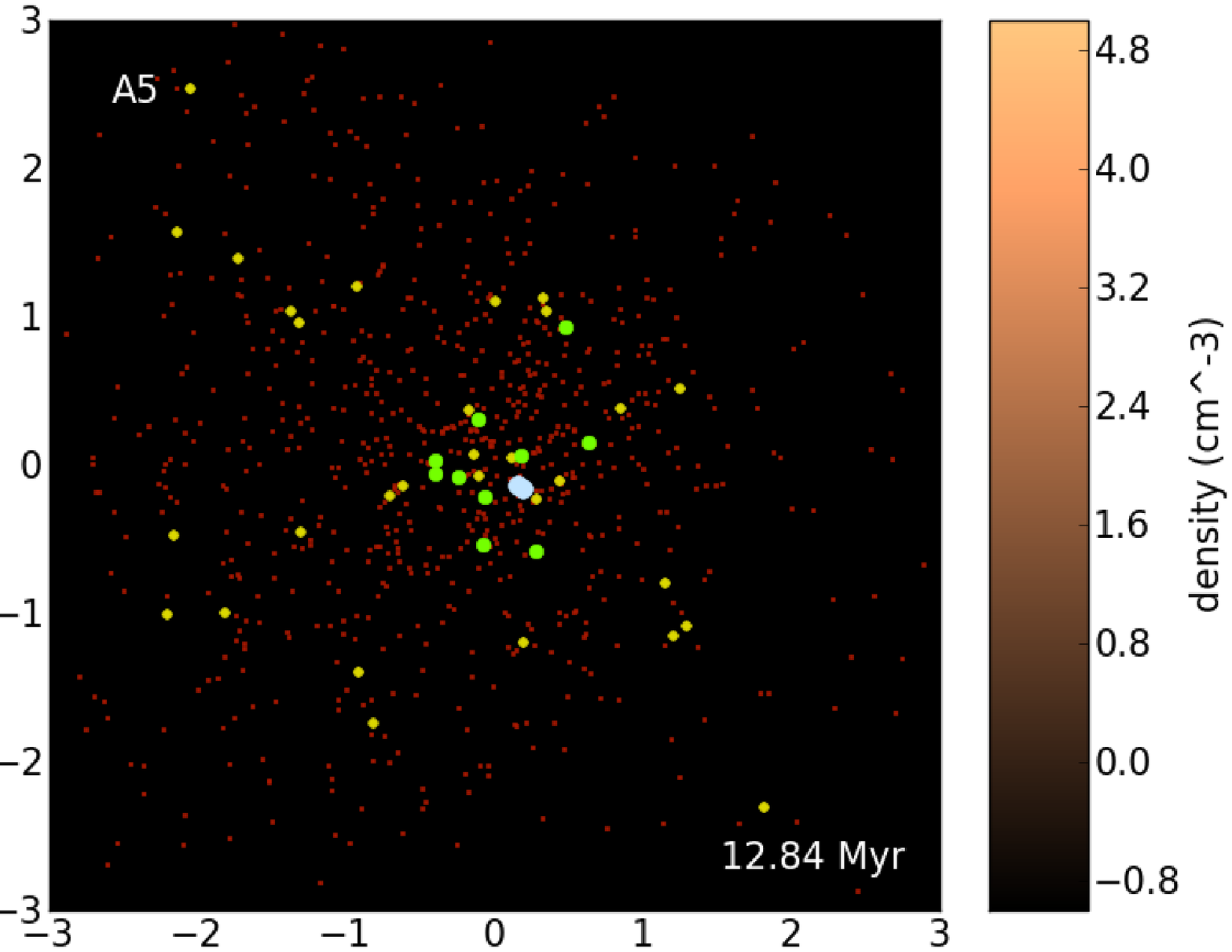,height=.3\textwidth}   %,angle=-90}

\caption{
  Stellar and gas distribution of the A2 and A5 runs.  The left panels
  show the gas and stellar distribution of the A2 run, those on the
  right panels the A5 run.  Snapshots are labeled by time in the lower
  right corner.  The density plots show cuts through the mid-plane.
  The points show stars in 4 mass groups ($m < 0.9$\,\Msun,
  $0.9\,\Msun < m< 2.5\,\Msun$, $2.5\,\Msun < m< 10\,\Msun$\, and $m>
  10\,\Msun$).
  \label{fig:model_A2}} 
\end{center}
\end{figure}

Figure\,\ref{fig:model_A2} shows the stellar and gas distribution of
our models A2 and run A5.  In both models we parameterize the relative
feedback efficiency between the stars and the gas by a parameter 
$f_{\rm fb}$, which is the fraction of the total supernova and wind 
energy output that ends up as thermal energy in the ISM (this accounts
for the uncertainties in modeling the feedback and radiative losses). 
For the A2 model we take $f_{\rm fb}=0.1$ while for the A5 model 
$f_{\rm fb}=0.01$.  The feedback is implemented by returning gas particles 
from stars in proportion to the mass loss rates of the stellar wind and SN,
with a thermal energy set by the mechanical luminosity of the star and 
the canonical energy $E_{\rm sn}=10^{51} \ergs$ in case a star 
goes supernova.  For each model we
plot the stars and a slice through the gas density distribution at
four moments during the simulation.  In the first A2 frame (at 0.96 Myr),
we see the early stages where stellar winds create buoyant bubbles
that rise out of the potential of the star cluster.  As the mechanical
luminosity increases these bubbles grow until they blow away sizable
fractions of the cluster gas and a free-flowing wind develops (4.37
Myr frame).  The strong feedback then unbinds most of the gas of the
cluster.  At approximately 9.5 Myr the cluster ISM has been
ejected---the gas visible in this frame originates from the strong AGB
wind of the most massive progenitor ($m\sim 21$\,\Msun).

At an age of 9.54 Myr the most massive star in the simulation
undergoes a supernova explosion which ejects the remaining gas from
the cluster (both simulations use the same initial realizations for
the IMF and stellar positions).  For the A5 simulation (with a
relative feedback efficiency of 0.01, compared to 0.1 for model A2),
the initial wind stages (before 0.96 Myr) proceed less violently,
with smaller bubbles, and a free-flowing wind does not
develop until just before the supernova (compare the 4.37 Myr
frames). The main difference between the A2 and A5 runs is that most
of the cluster gas is retained in the latter case until the first
supernova blows it away. Just before the supernova the A5 cluster is
much more compact than in the A2 run.

\begin{figure}[htbp]
\begin{center}
\psfig{figure=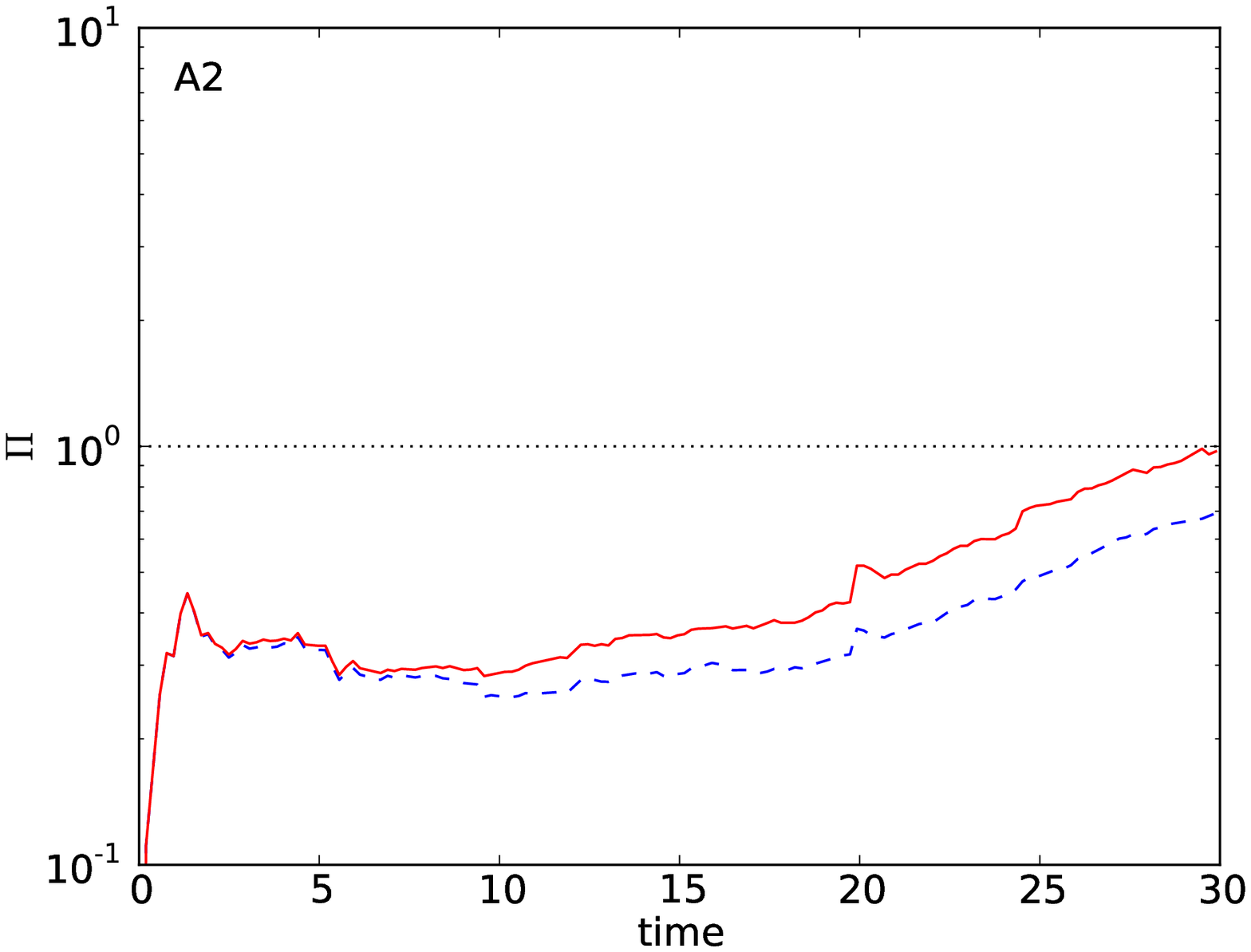,width=0.49\columnwidth} %,angle=-90} 
~\psfig{figure=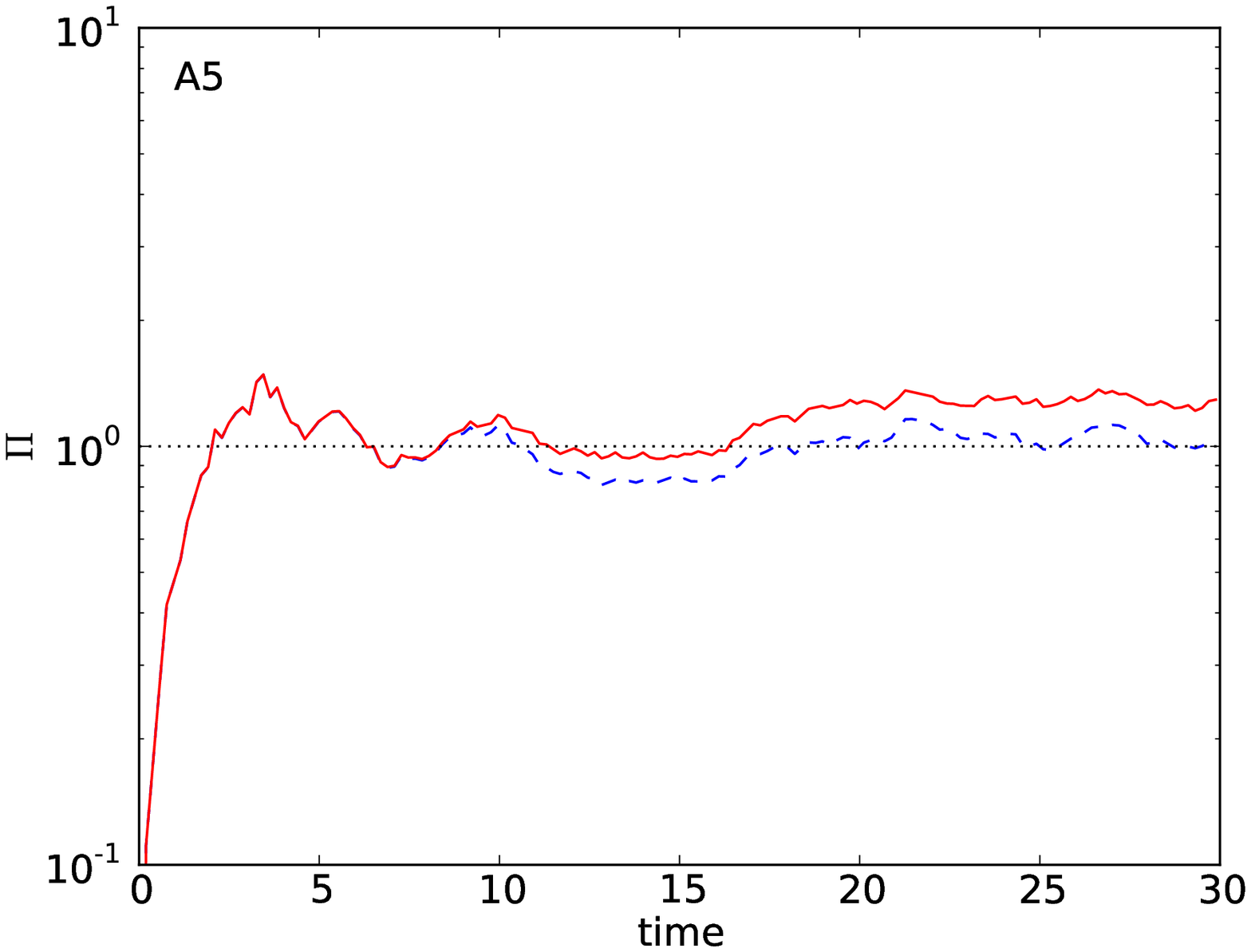,width=0.49\columnwidth}   %,angle=-90} 
\caption{ The ratio of the cluster age to the calculated
  crossing time as defined by \cite{2011MNRAS.410L...6G} ($\Pi \equiv$
  age/\Tcr) for models A2 and A5 (see also Fig.\,\ref{fig:model_A2}).
  \label{fig:model_PI}}
\end{center}
\end{figure}

Loosely bound associations can be distinguished from `true' stellar clusters 
using limited observables by considering the ratio $\Pi$ of age $T_{\rm cl}$ 
and the crossing time $T_{\rm cross}$ \cite{2011MNRAS.410L...6G}:
\begin{equation}
\label{eq_pi}
\Pi \equiv T_{\rm cl}/T_{\rm cross}
\end{equation}
In Fig.\,\ref{fig:model_PI} we present the value of $\Pi$ for models
A2 and A5 as functions of time. A value of
$\Pi<1$ indicates that the cluster is in a ballistic state of
expansion, whereas a value of $\Pi>1$ implies a bound state. 
In Fig.\,\ref{fig:model_PI} we see two completely
different behaviors for the evolution of $\Pi$.  For model A2, $\Pi$
rises sharply but the cluser fails to reach a bound
state (in the sense of $\Pi>1$) until much later
(after $t \apgt 30\,Myr$), whereas model A5 reached a bound
state within a few Myr after formation and remained marginally bound
for the stof the simulation (up to $\sim 30$\,Myr).  In our 
survey of parameter space (see Pelupessy \& Portegies Zwart 2011), 
we explore a wider range of initial conditions. 
% Currently
% it is  hard to explain the high values of {$\Pi$} within the
% first $\sim 5$\,Myr of the cluster lifetime.

\section{Conclusions}

We have simulated star clusters in their embedded phase.  Our
simulations include the gravitational dynamics of the stars, the
dynamics of the intracluster gas, and the internal evolution of the
stars.  We find that the star formation efficiency is a poor predictor
of final state of the cluster.  There are several arguments why the
star formation efficiency is less important than has been found in
earlier studies.  The most dramatic event in
the lifetime of a young cluster is the occurrence of the first
supernova, which blows away most of the residual gas in the cluster.
But due to earlier fast Wolf-Rayet winds from the massive stars most of the
gas has already escaped without much damage to the cluster. In addition, during
the time between the strong Wolf-Rayet wind and the supernova
explosion the cluster has time to relax, making it more resilient
against destruction by the loss of primordial gas. 

Statistical variations in our method of generating the initial mass
function have a profound effect on the early evolution of the cluster.
The survival of the star cluster may well depend on the masses and orbits
of the few most massive stars it contains. A cluster with a slight
enhancement of massive stars may well dissolve, whereas a more
fortunate cluster may be born with a larger gap between the masses of
few most massive stars.  Slight differences of even a few {\Msun} in
the most massive stars may well be crucial in determining the survival
of the cluster.

The surviving clusters are strongly mass segregated. During the
embedded phase massive stars easily sink to the cluster center. The
degree of mass segregation found in the surviving clusters nicely
matches those required to explain the oberved degree of mass
segregation in the Pleiades.

Our prescription for the radiative feedback in our models is still
very limited, and the next obvious step in improving our model would
be by adopting a radiative transfer code to resolve this problem.

\section*{Acknowledgments}

This work was supported by NWO (grants \#643.200.503, \#639.073.803
and \#614.061.608), NOVA and the LKBF in the Netherlands, and by NSF
grant AST-0708299 in the U.S.

\end{document}